\documentclass[twocolumn,english,prl,floatfix]{revtex4}
\usepackage[T1]{fontenc}
\usepackage[latin9]{inputenc}
\usepackage{amssymb}

\makeatletter
\@ifundefined{textcolor}{}
{%
 \definecolor{BLACK}{gray}{0}
 \definecolor{WHITE}{gray}{1}
 \definecolor{RED}{rgb}{1,0,0}
 \definecolor{GREEN}{rgb}{0,1,0}
 \definecolor{BLUE}{rgb}{0,0,1}
 \definecolor{CYAN}{cmyk}{1,0,0,0}
 \definecolor{MAGENTA}{cmyk}{0,1,0,0}
 \definecolor{YELLOW}{cmyk}{0,0,1,0}
 }

\usepackage{graphicx}

\makeatother

\usepackage{babel}

\begin{document}

\title{An exact lower energy bound for the infinite square well potential}

\author{M. Ögren$^{1}$, M. Carlsson$^{2}$}

\affiliation{$^{1}$ARC Centre of Excellence for Quantum-Atom Optics, School of
Mathematics and Physics, University of Queensland, Brisbane, Queensland
4072, Australia \\
 $^{2}$Mathematics Department, Purdue University, 150 N. University
St., West Lafayette, 47907 IN, U.S.}

\date{\today{}}
\begin{abstract}
We give a lower bound for the energy of a quantum particle in the
infinite square well. We show that the bound is exact and identify
the well known element that fulfils the equality. Our approach is
not direct dependent on the Schrödinger equation and illustrates an
example where the wavefunction is obtained direct by energy minimization.
The derivation presented can serve as an example of a variational
method in an undergraduate level university course in quantum mechanics.
\end{abstract}
\maketitle

\section{Introduction}

In the introductory study of quantum mechanics the three probably
most basic examples to master for the student are: the infinite square
well, the harmonic oscillator and the (non-relativistic) Hydrogen
atom, see e.g. \cite{QMbook}. These can all be solved exactly from
the Schrödinger equation with moderate knowledge in mathematics. Their
solutions illustrate the emergence of (bound) quantized states, hence
a dramatic deviation from the classical picture. The energy spectrum
for these three integrable examples can also be solved by semiclassical
quantization methods \cite{Keller}. Remarkably all exact energies
can be obtained analytically, relying on the fact that their Hamilton
functions are quadratic forms (after a coordinate transformation in
the Hydrogen case \cite{Kustaanheimo}). These simple concepts are
still very important ingredients in today's research frontier in (many-body)
quantum physics, see e.g. \cite{Pricoupenko}, \cite{Ogren} and \cite{Lukin}
respectively, for a heuristic model of a Fermi gas at unitarity, too
see the emergence of super-shell structures in a harmonic trap and
to model quantum information processing with Rydberg atoms.

There are not so many other examples of potentials that allow for
an exact analytic treatment. For example the finite square well is
of great practical importance and it helps in understanding the physics
of modern man made low-dimensional structures and quantum well devices.
However, it can not be solved exact although many practical (some
also analytic) approximations exists \cite{Bonfim}. In the following
we discuss the obviously more artificial infinite square well, while
the present notes does not provide any new results, it highlights
an alternative path \cite{Goodman}, and can be used as a simple example
of a variational method.

\section{The infinite square well }

The infinite square well potential is here defined as

\begin{equation}
V\left(x\right)=\left\{ \begin{array}{l}
0\:\:{\it if}\:\:0\leq x\leq L\\
\infty\:\: else\end{array}\right.,\label{wellpotential}\end{equation}
where the wavefunctions to the potential of Eq. (\ref{wellpotential})
obey 

\begin{equation}
\psi\left(0\right)=\psi\left(L\right)=0.\label{boundaries}\end{equation}
This agrees with the probability interpretation, that the particle
can not be found in finite regions where $V=\infty$. The usual way
to obtain the lowest energy (eigenvalue) is to solve the following
Schrödinger equation

\begin{equation}
-\frac{\hbar^{2}}{2m}\frac{d^{2}\psi}{dx^{2}}=E\psi,\:\:0\leq x\leq L.\label{SE}\end{equation}
The point of the current note, however, is to show how to accurately
estimate the lowest energy $\min\left(E\right)$ without explicit
use of the Schrödinger equation or its eigenvalues.

\section{Heisenberg inequality}

A standard approach is to start from the Heisenberg uncertainty relation
for the momentum and position \cite{Kennard}

\begin{equation}
\Delta p\Delta x\geq\frac{\hbar}{2}.\label{eq:HeisenbergsRelation}\end{equation}
This gives an approximate bound for the groundstate energy \cite{Yue}
in the potential of Eq. (\ref{wellpotential}), where $\Delta x\sim L/2$ 

\begin{equation}
E=\frac{\left\langle \hat{p}^{2}\right\rangle }{2m}\sim\frac{\left(\Delta p\right)^{2}}{2m}\geq\frac{\hbar^{2}}{2mL^{2}}.\label{EfromHeisenberg}\end{equation}
Since the energy spectrum to Eq. (\ref{SE}) is 

\begin{equation}
E_{n}=\frac{\hbar^{2}\pi^{2}n^{2}}{2mL^{2}}\:,\: n=1,2,...\:\Rightarrow\:\min\left(E\right)=E_{1}.\label{En}\end{equation}
The lower bound of the estimate in Eq. (\ref{EfromHeisenberg}) is
hence predicting the true lowest energy $E_{1}$ wrong by a factor
$\pi^{2}$ . This is only acceptable for very crude estimates \cite{RefinedAnalysis},
such as comparing the energy of systems of different sizes $L$ (e.g.
nuclei {[}\emph{MeV}{]} and atoms {[}\emph{eV}{]}).

\section{An exact inequality}

We now derive a stricter lower bound of the groundstate energy to
the infinite square well potential defined in Eq. (\ref{wellpotential}).
Let us denote the wavefunction of the groundstate with $\psi_{0}$,
this should obey the boundary conditions of Eq. (\ref{boundaries})
together with the normalization condition 

\begin{equation}
\int_{0}^{L}\left|\psi\left(x\right)\right|^{2}dx=1.\label{normalisation}\end{equation}
Under those subsidiary conditions the groundstate $\psi_{0}$ should
minimize the energy functional (here $\hat{H}=\left|\hat{p}\right|^{2}/2m$,
where we use $\hat{p}=-i\hbar\, d/dx\,$ \cite{Levich} and integration
by parts)

\begin{equation}
E\left(\psi\right)=\left\langle \psi\left|\hat{H}\right|\psi\right\rangle =\frac{\hbar^{2}}{2m}\int_{0}^{L}\left|\frac{d\psi\left(x\right)}{dx}\right|^{2}dx.\label{energyfunctional}\end{equation}
Our main task here is to prove the following inequality

\begin{equation}
E\left(\psi\right)\geq\frac{\hbar^{2}\pi^{2}}{2mL^{2}},\label{inequality}\end{equation}
without using the Schrödinger equation (\ref{SE}), which is the Euler-Lagrange
equation to (\ref{energyfunctional}). We then find the element $\psi_{0}$
which gives equality in Eq. (\ref{inequality}). Combining Eqs. (\ref{normalisation})
and (\ref{energyfunctional}) we can write the inequality of Eq. (\ref{inequality}) 

\begin{equation}
\int_{0}^{L}\left|\frac{d\psi\left(x\right)}{dx}\right|^{2}dx\geq\frac{\pi^{2}}{L^{2}}\int_{0}^{L}\left|\psi\left(x\right)\right|^{2}dx.\label{inequalitytoprove}\end{equation}
We remark that a related inequality have been proved with geometrical
methods by Wilhelm Wirtinger already in the 19th century, see e.g.
\cite{wirtinger}.

\section{Proof of the inequality}

We expand $\psi$ to an odd function $\tilde{\psi}$ on the interval
$-L\leq x\leq L$, hence $\tilde{\psi}\left(-L\right)=\tilde{\psi}\left(L\right)=0$
{[}Eq. (\ref{boundaries}){]}. It is natural to express $\tilde{\psi}$
as a Fourier series $\tilde{\psi}=\sum_{k}a_{k}\phi_{k}$, with $\phi_{k}=\exp\left(\pi ikx/L\right)/\sqrt{2L}$.
It then follows that

\begin{equation}
\frac{d\tilde{\psi}}{dx}=\sum_{k}\frac{\pi ik}{L}a_{k}\phi_{k},\label{eq:derivative}\end{equation}
(using that $d\psi/dx$ is a continuous function in $0<x<L$, that
$\tilde{\psi}\left(-L\right)=\tilde{\psi}\left(L\right)$ and integration
by parts.) This transforms Eq. (\ref{inequalitytoprove}) into 

\begin{equation}
\int_{-L}^{L}\Bigl(|\sum_{k\neq0}ka_{k}\phi_{k}|^{2}-|\sum_{k\neq0}a_{k}\phi_{k}|^{2}\Bigr)dx\geq0,\label{eq:ProofI}\end{equation}
since $a_{0}=\int_{-L}^{L}\tilde{\psi}\phi_{0}dx=0$ as $\tilde{\psi}$
is odd. Using the orthogonality of the $\phi_{k}$:s gives

\begin{equation}
\sum_{k\neq0}\left(k^{2}-1\right)\left|a_{k}\right|^{2}\geq0,\label{eq:ProofII}\end{equation}
and hence the inequality ($\geq$) is proved. The equality ($=$)
is seen to be fulfilled for the elements $\psi_{0}$ such that $a_{-1}=-a_{1}\neq0$
(since $\tilde{\psi}$ is odd) and $a_{k}=0$ for all other $k$.
Since we choose $\psi_{0}$ to be positive and real, this means that
the normalized wavefunction on $0\leq x\leq L$ which minimize the
energy is

\begin{equation}
\psi_{0}\left(x\right)=e^{3\pi i/2}\left(\phi_{1}-\phi_{-1}\right)=\sqrt{2/L}\sin\left(\pi x/L\right),\label{eq:psi_0}\end{equation}
as desired. The groundstate momentum relation $p=\hbar\pi/L=\sqrt{2mE}$
then also follows from Eq. (\ref{inequality}). We finally remark
that other basis can be used to prove Eq. (\ref{inequalitytoprove})
by expanding $\psi$, but the equality with Eq. (\ref{eq:psi_0})
then in general have to be checked by projection.

\section{Conclusions}

We have obtained the well known groundstate of the infinite square
well analytically without directly solving the Schrödinger equation.
This derivation can serve as a neat pedagogical tool in undergraduate
quantum mechanics classes. It stresses the view that the shape of
the wavefunction is such that the energy is minimized, which is widely
used for approximations to more complicated (many-body) systems.


\vspace{-5mm}
\begin{thebibliography}{13}
\bibitem{QMbook} B. H. Bransden and C. J. Joachain, Quantum Mechanics,
2.nd ed. Englewood Cliffs, NJ: Prentice Hall (2000). 

\bibitem{Keller} J. B. Keller, Ann. Phys. {\bf 4}, 180 (1958).

\bibitem{Kustaanheimo} P. Kustaanheimo and E. Stiefel, J. Reine Angew.
Math. {\bf 218}, 204 (1965).

\bibitem{Pricoupenko} L. Pricoupenko and Y. Castin, Phys. Rev. A
{\bf 69}, 051601(R) (2004).

\bibitem{Ogren} M. Ögren and H. Heiselberg, Phys. Rev. A {\bf 76},
021601(R) (2007).

\bibitem{Lukin} M. D. Lukin \emph{et al.}, Phys. Rev. Lett. {\bf 87},
037901 (2001).

\bibitem{Bonfim} O. F. de Alcantara Bonfim and D. J. Griffiths, Am.
J. Phys. {\bf 74}, 43-48 (2005).

\bibitem{Goodman} M. Goodman, Am. J. Phys. {\bf 49}, 843-847 (1981).

\bibitem{Kennard} E. H. Kennard, Zeitschrift für Physik {\bf 44},
326 (1927).

\bibitem{Yue} Z. Yue, Am. J. Phys. {\bf 58}, 554-556 (1990).

\bibitem{RefinedAnalysis} A refined analysis gives $\Delta p\Delta x=\sqrt{\pi^{2}/12-1/2}\,\hbar\approx0.57\hbar>\hbar/2$
for the infinite square well. For a discussion of the convergence
of the moments $\left\langle \hat{p}^{s}\right\rangle $, see e.g.
F. E. Cummings, Am. J. Phys. {\bf 45}, 158-160 (1977).

\bibitem{Levich} B. G. Levich, Y. A. Vdovin and V. A. Myamlin, Theoretical
physics 3, an advanced text, Quantum mechanics, vol. 3, North-Holland
Publishing Company 1973.

\bibitem{wirtinger} A. Pressly, Elementary differential geometry,
Springer-Verlag London 2002.
\end{thebibliography}
\end{document}